\documentclass[prl,showpacs,twocolumn]{revtex4}%
\usepackage{amsmath}
\usepackage{amsmath}
\usepackage{graphicx}
\usepackage{amsfonts}
\usepackage{amssymb}%
\begin{document}
\title{Theory of dark-state polariton collapses and revivals}
\date{\today}
\author{S. D. Jenkins}
\author{D. N. Matsukevich}
\author{T. Chaneli\`{e}re}
\author{A. Kuzmich}
\author{T. A. B. Kennedy}
\affiliation{School of Physics, Georgia Institute of Technology, Atlanta, Georgia 30332-0430} 
\begin{abstract}
We investigate the dynamics of dark-state polaritons in an atomic
ensemble with ground-state degeneracy. A signal light pulse may be
stored and retrieved from the atomic sample by adiabatic variation
of the amplitude of a control field. During the storage process, a
magnetic field causes rotation of the atomic hyperfine coherences,
leading to collapses and revivals of the dark-state polariton
number. These collapses and revivals should be observable in
measurements of the retrieved signal field, as a function of
storage time and magnetic field orientation.
\end{abstract}
 \maketitle

A quantum memory element consisting of an ensemble of atoms, with
efficient coupling to a signal light field, represents a node in
several quantum network architectures
\cite{fleischhauer,duan,lukin,saffman}.
A dark-state polariton (DSP) is a collective excitation, with light
field and atomic spin wave parts, in which the relative size of the
light and matter contributions can be varied by changing the amplitude
of a control laser field \cite{fleischhauer}.
In connection with atomic memories, DSPs offer the possibility for
efficient transfer of information between a light carrier and an atomic
medium, with programmable storage of the excitation in the atomic spin
coherence.
The storage and subsequent retrieval of the signal field component of
the DSP can be achieved by the extinction and subsequent
reactivation of the control field after a given storage time.
Experimental demonstrations of ``stopped-light" can be understood in
terms of the concept of DSP in just this way
\cite{phillips,hau,mair}.

In a recent work the storage and retrieval of single photons using
an atomic ensemble based quantum memory was reported, with a
storage time conjectured to be limited by inhomogeneous broadening
in the ambient magnetic field \cite{chaneliere}. During the
storage, the DSP consists entirely of atomic spin-wave, and in
order to understand its dynamics in a magnetic field it is
necessary to properly account for the atomic level degeneracy and
the signal and control field polarizations. In particular for
alkali atoms, which have non-zero nuclear spin, the electronic
levels have hyperfine structure. In this case we must define a
more general form of DSP field operator than that
of a simple lambda configuration, in which the atomic spin-wave
part corresponds to a particular superposition of hyperfine
coherences of the ground electronic level. These coherences are,
in turn, intimately related to the phenomenon of
electromagnetically-induced transparency (EIT)
\cite{harris,scully,LingXiaoPRA1996}. We shall see that in a magnetic field the
temporal evolution of the DSP reveals a series of collapses and
revivals due to the evolution of its spin-wave component during
the storage phase of the process. We predict that the collapses
and revivals should be directly observable in measurements of the
retrieved signal field as a function of storage time.

\begin{figure}
  \includegraphics[width=8.4cm]{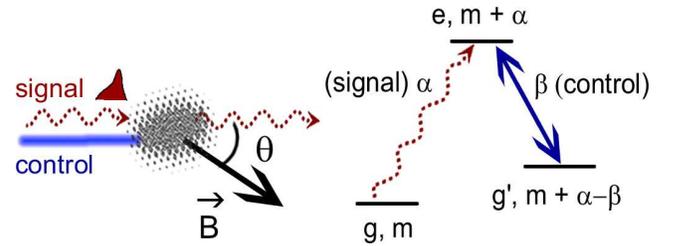}
  \caption{(Color online) On the left, a diagram shows an atomic ensemble
  interacting with
  copropagating signal and control fields.
  The signal (helicity $\alpha$), resonant on the $g \leftrightarrow
  e$ transition, is stored and subsequently retrieved by variation of a
  control field (helicity $\beta$), which resonantly couples levels $g'$
  and $e$.  A constant magnetic field $\mathbf{B}$, oriented at
  an angle $\theta$ from the propagation axis, rotates the atomic
  coherences during the storage.
  For each state $|g,m\rangle$ in level $g$, there is an associated
  $\Lambda$ configuration, as shown on the right.
  The signal
  connects the states $|g,m\rangle$ and $|e,m+\alpha\rangle$, while
  the control field drives transitions between
  $|e,m+\alpha\rangle$ and $|g',m+\alpha-\beta \rangle$.  }
  \label{Fig1}
\end{figure}
We develop the theory of EIT in a degenerate atomic medium with ground
levels $g$ and $g^{\prime}$ and excited level $e$, which have energies
$\hbar \omega_g \equiv 0$, $\hbar \omega_{g'}$ and $\hbar \omega_e$,
respectively (Fig. \ref{Fig1}).
The Zeeman states of level $g$ are written $|g,m\rangle$, where $-F_g
\le m \le F_g$; similar definitions hold for the other levels.
All $N$ atoms are assumed to be initially prepared in level $g$
without polarization, i.e., the density matrix of atom $\mu$ is
$\hat{\rho}_\mu = \sum_m p \vert g,m\rangle_{\mu}\langle g,m \vert$
where we write $p=1/(2F_g+1)$.
The density of atoms in the sample is assumed to be constant.
The atoms experience a uniform magnetic field $\mathbf{B}$ oriented at
an angle $\theta$ with respect to the light propagation $z$ axis.
The magnetic field-atom interaction $\hat{V}_B =
\mu_B\mathbf{B}\cdot\sum_{s=g,g^{\prime},e} g_s\mathbf{\hat{F}_s}$,
where $\mathbf{\hat{F}}_s$ is the projection of the atomic angular
momentum operator onto level $s$ and $g_s$ is the corresponding
Land\'{e} $g$ factor.
The magnetic field induces a Larmor spin precession which is primarily
important in the storage phase, when the signal field amplitude is
zero.
We note that in prior work collapses and revivals of single-atom
Zeeman coherences have been observed \cite{pritchard,smith}.

We proceed by generalizing the treatment of Fleischhauer and Lukin \cite{fleischhauer1} to include the degenerate atomic level scheme and the
presence of a magnetic field. We assume the number of photons contained in the signal pulse is much less than the number of atoms in the sample.
For an initially unpolarized sample in level $g$, the ground state populations and \emph{Zeeman} coherences, as opposed to hyperfine coherences,
are unaffected by the signal, control and magnetic fields. Our treatment can be extended to an initially spin-polarized atomic sample, as we will
report in a separate publication. The signal field, which we assume propagates in the positive $z$ direction, and atomic coherence operators
satisfy the quantum Langevin equations
\begin{subequations}
\label{EqLangevin}
\begin{equation}
  \left(  \frac{\partial}{\partial t}+c\frac{\partial}{\partial z}\right)
          \hat{\Phi}_{\alpha} = iN \kappa^{\ast}
      \sum_{m=-F_{g}}^{F_{g}} C_{m\alpha} \hat{Q}_{e~m+\alpha}^{g~m}
\end{equation}%
\begin{equation}
  \frac{\partial}{\partial t}\hat{Q}_{g^{\prime}~m^{\prime}}^{g~m}
  =i\Omega(t)
     C_{m^{\prime}\beta}^{\prime}
     \hat{Q}_{e~m^{\prime}+\beta}^{g~m}
\end{equation}
\begin{eqnarray}
&&  \left(\frac{\partial}{\partial{}t} + \frac{1}{2}\Gamma_e\right)
     \hat{Q}_{e~m^{\prime}}^{g~m}
   - \hat{F}_{e~m^{\prime}}^{g~m} \nonumber\\
&& = i\Omega(t)
        C_{m^{\prime}-\beta,\beta}^{\prime}
        \hat{Q}_{g^{\prime}~m^{\prime}-\beta}^{g~m}
   +  i\delta_{m+\alpha,m'}\kappa pC_{m\alpha}%
      \hat{\Phi}_{\alpha}
\end{eqnarray}
\end{subequations}
where the slowly varying electric field of helicity $\alpha=\pm1$ is given by $\Phi_\alpha(z,t)\equiv i \sum_k \hat{a}_{k,\alpha}\exp(i(qz +
\omega_e t))$, $q=k-(\omega_e/c)$ and the collective atomic coherence is defined $\hat{Q}_{s^{\prime}~m^{\prime}}^{s~m}(z,t) \equiv (1/N_z)
\sum_{\mu=1}^{N_z} \hat{\sigma}_{s\,m,\,s^{\prime}\,m^{\prime}}^{(\mu)} \exp(-i(\omega_{s}-\omega_{s^{\prime}}) (t-z/c) )$ \cite{haake}, where
$\mu=1\ldots N_z$, $\hat{\sigma}_{s\,m,\,s^{\prime}\,m^{\prime}}^{(\mu)}$ is the $\mu$'th atom hyperfine coherence operator, and $N_z = N dz/L$
is the number of atoms between $z$ and $z+dz$; $L$ is the length of the sample. The control field is assumed to have the circular polarization
$\beta=\pm 1$ and real Rabi frequency $\Omega(t)$, which is a specified function of time, and $\kappa$ is the coupling constant for the probe
transition. We adopt the shorthand for the Clebsch-Gordan coefficients $C_{m\alpha} \equiv C_{m~\alpha~m+\alpha}^{F_g~1~F_e}$ and
$C^{\prime}_{m\beta} \equiv C_{m~\beta~m+\beta}^{F_{g^{\prime}}~1~F_e}$; it is useful to define
$R_{m\alpha}(\beta)=C_{m\alpha}/C^{\prime}_{m+\alpha-\beta,\beta}$. The decay rate of level $e$ is denoted by $\Gamma_e$ and
$\hat{F}_{e~m'}^{g~m}$ is a corresponding quantum noise operator. The coupling of the atoms to the uniform magnetic field can be taken into
account by the addition of appropriate commutators with the interaction $\hat{V}_B$ in the atomic equations.

We first establish some standard features of EIT with our model.
The propagation of a classical (coherent) signal through the medium is
found by dropping the quantum noise operator, and replacing the field
and coherence operators with their respective expectation values.
For a constant amplitude control field, the linear susceptibility for
the signal field of angular frequency $\omega$ is found to be
\begin{equation}
  \label{EqSusceptability}
  \chi_\alpha(\Delta) \approx \frac{cd_\alpha}{2\omega_eL}
          \sum_m \frac{\Gamma_e \Delta X_{m\alpha}^2
                       (\Omega^2 C_{m+\alpha-\beta,\beta}^{\prime 2} - \Delta^2
                        + i\Delta\Gamma_e/2)}
                      {\left(\Omega^2 C_{m+\alpha-\beta,\beta}^{\prime2}
                             - \Delta^2\right)^2 +
                             (\Delta\Gamma_e/2)^2},
                             \end{equation}
where $\Delta \equiv \omega-\omega_e$ is the detuning of the
signal from atomic resonance, and $X_{m\alpha} \equiv
C_{m\alpha}/\sqrt{\sum_{m^{\prime}}C_{m^{\prime}\alpha}^2}$.
The dimensionless quantity $d_\alpha$ is the optical thickness, which
is defined such that $\exp(-d_\alpha)$ is the on-resonance intensity
transmittance in the absence of a control field, and can be expressed as
\begin{equation}
  \label{EqOptThickness}
  d_\alpha = 6\pi\eta \frac{N}{A}\left(\frac{c}{\omega_e}\right)^2 p
         \sum_m C_{m\alpha}^2
\end{equation}
where $\eta$ is the fraction of atoms in excited level $e$ that spontaneously decay into ground level $g$, and $A$ is the cross sectional area of
the ensemble. When a control field is present, an EIT window exists provided that the Clebsch-Gordan coefficients
$C_{m+\alpha-\beta,\beta}^{\prime}$ do not vanish for any $-F_g \le m \le F_g$ for which $C_{m\alpha}\ne0$. If, however,
$C_{m+\alpha-\beta,\beta}^{\prime}=0$,  and $C_{m\alpha} \ne 0$, it means that there is an excited state $\vert e,m+\alpha \rangle$ not coupled
by the control field to a state in the ground level $g^{\prime}$, i.e., there is an unconnected lambda configuration. The subset of atoms
initially in the state $\vert g,m \rangle$ would absorb the signal field and spontaneously emit radiation as if there were no control field
present. In order for EIT to exist, one must make a judicious choice of atomic levels and signal and control field polarizations.

Assuming a choice of polarizations that supports EIT, we are able
to generalize the adiabatic treatment of Ref. \cite{fleischhauer}
to Eq.~(\ref{EqLangevin}) to derive the DSP operator for helicity
$\alpha$, with control field polarization $\beta$
\begin{equation}
  \hat{\Psi}_{\alpha}\left(z,t\right)
  = \frac{\Omega(t) \hat{\Phi}_{\alpha}( z, t)
          - N \kappa ^{\ast}%
          \sum_{m} R_{m\alpha}(\beta)
      \hat{Q}_{g^{\prime}~m+\alpha-\beta}^{g~m} ( z,t)}
  {\sqrt{ \Omega(t)^{2}
         + Np\left\vert \kappa \right\vert ^{2}
           \sum_{m}
       \left\vert R_{m\alpha}(\beta)\right\vert ^{2}}}.%
\end{equation}
As in Ref.\cite{fleischhauer}, this operator obeys the simple
propagation equation $(\partial/\partial{}t + v_g\partial/\partial
z)\hat{\Psi}_\alpha(z,t)=0$ with the reduced group velocity
$v_g=c\Omega^2/(\Omega^2+Np\vert \kappa \vert^2\sum_m\vert
R_{m\alpha}(\beta)\vert^2)$ which can be adiabatically controlled by
time dependent variation of $\Omega(t)$.
From the definition of $\hat{\Psi}$, we see that as $\Omega$ goes to
zero, the wave excitation stops propagating and transforms into a
particular linear combination of hyperfine coherences $\sim \sum_{m}
R_{m\alpha}(\beta) \hat{Q}_{g^{\prime}~m+\alpha-\beta}^{g~m} ( z,t)$.
This nontrivial result arises from the treatment of the full
degeneracy of the atomic ensemble; only this combination of hyperfine
coherences is adiabatically transformed into the signal field
\emph{via} the control field retrieval process.
Orthogonal combinations of hyperfine coherences couple to optical
coherences in the presence of the control field and result in excited
state spontaneous emission; we will refer to these as the bright-state
polariton (BSP) component.
It is also possible that some population of atoms remains trapped in
the ground states, and is unaffected by the control field.

In order to demonstrate the importance of the dark state polariton in
the signal storage and retrieval process in a magnetic field, we
numerically solve Eqs. (\ref{EqLangevin}) for a coherent signal field.
We thus calculate the expectation values of the spin wave coherences
$\langle \hat{Q}_{g^{\prime} m+\alpha-\beta}^{g m}(z,t)\rangle$ and
the signal field, allowing us to determine the DSP and BSP
components.
This is accomplished by defining a vector space of $2F_g+2$
dimensions, with orthonormal basis vectors $\mathbf{e}_m$, each
corresponding to a hyperfine coherence $\hat{Q}_{g^{\prime}
  m+\alpha-\beta}^{g m}$, and $\mathbf{e}_{\Phi}$ corresponding to the
signal field.
We define the coherence ``vector'' $\mathbf{v}\equiv\left\langle
\hat{\Phi}_{\alpha} \right\rangle \mathbf{e}_{\Phi}+\sum
_{m}\sqrt{N/p}\left\langle
\hat{Q}_{g^{\prime}~m+\alpha-\beta}^{g~m}\left(  z,t\right)
\right\rangle \mathbf{e}_{m}$.
We note that this is not normalized since its magnitude is dependent
on both the time dependent signal and control fields.
Associated with the DSP we define a vector
$\mathbf{u}_{\Psi}\equiv\Omega\mathbf{e}_{\Phi}+\sqrt{Np}\kappa
\sum_{m} R_{m\alpha}(\beta)\mathbf{e}_{m}$ and the corresponding unit
vector $\mathbf{e}_{\Psi} = \mathbf{u}_{\Psi}/
||\mathbf{u}_{\Psi}||$.
We then determine the DSP component $p_D = |\mathbf{e}_{\Psi} \cdot
\mathbf{v}|^2$, and BSP component $p_B = ||\mathbf{v} -
\mathbf{e}_{\Psi}\mathbf{e}_{\Psi} \cdot \mathbf{v}||^2$.

\begin{figure}
  \includegraphics{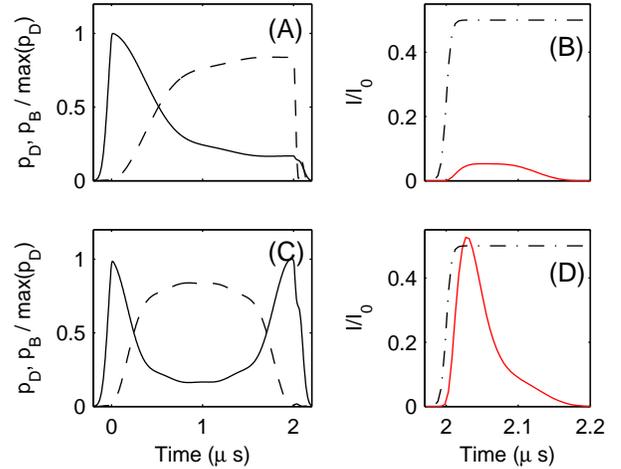}
  \caption{(Color online) Numerical results illustrate the storage and retrieval of
    a signal pulse from an atomic ensemble as described in the text.
    We show results for two values of the magnetic field oriented
    along the $z$ axis.  Panels (A) and (B) correspond to a magnetic
    field $B=0.267~\mbox{G}$, so that the signal is stored for
    $T_{L}/4$, where $T_L=8 \operatorname{\mu s}$ is the Larmor period.  Panels (C) and (D) show results
    for $B=0.535~\mbox{G}$, corresponding to a signal storage time of
    $T_L/2$, where $T_{L}=4\operatorname{\mu s}$.
    The signal field intensity transmittance
    $I(t)/I_0$ (Solid
    line) and control field Rabi frequency (dot-dashed line), displayed in arbitrary
    units,
    are shown in
    panels (B) and (D).  Panels (A) and (C) display scaled dark
    state $p_D$ (solid line) and bright state $p_B$ (dashed line)
    polariton components, as explained in the text.  In panel (B) the ratio of
    retrieved to input signal pulse energy is $4.38 \%$ while in (D)
    the ratio is $25.09 \%$. }
  \label{FigDSPolariton}
\end{figure}

As an example of signal storage and retrieval we consider an atomic
sample of $^{85}$Rb, in which the control field and signal field
polarizations are chosen equal $\alpha=\beta=1$, and the optical
thickness $d=8$, Fig.~\ref{FigDSPolariton}.
The atomic levels $g,g'$ and $e$ correspond to the $5S_{1/2}$
$F={2,3}$ and $5P_{1/2}$ $F={3}$ levels of the $D_1$ line,
respectively.
The spontaneous decay rate $\Gamma_{e}/(2\pi) =
5.98\operatorname{MHz}$.
The incident signal field has a Gaussian envelope of full width half
maximum $120\operatorname{ns}$, and the peak enters the
$3\operatorname{mm}$ long sample at $t=-60 \operatorname{ns}$.
The control field has a constant Rabi frequency $\Omega = 1.5\Gamma_e$
until it is smoothly turned off at $t=0$ over a period of $20
\operatorname{ns}$, when a fraction of the signal field is converted
into hyperfine coherences of the atomic spin wave.
%%%%%% I don't particularly like this one
The excitation is  stored from $0 \le t \le 2 \operatorname{\mu s}$ in the presence of the magnetic field, before the control field is
reactivated, and the signal field retrieved. In Fig.~\ref{FigDSPolariton}, panels (A) and (B), the magnetic field is chosen so that the storage
time corresponds to a quarter of a Larmor period $T_L \equiv 2\pi\hbar/\vert g_g\mu_B \mathbf{B}\vert$, while in panels (C) and (D), the storage
time is $T_L/2$. In panel (A) $p_D$ grows as the signal pulse arrives at the point of observation, and reaches a peak when the control field is
switched off. It then decays during the storage phase, due to Larmor precession of the hyperfine coherences in the applied magnetic field, which
causes the corresponding growth of $p_B$. When the control field is reactivated, $p_B$ decays rapidly due excited level coupling and subsequent
spontaneous emission, though $p_D$ remains finite as the spin wave coherence of the DSP is converted into the forward propagating signal field;
the retrieved signal field intensity is illustrated in panel(B). In panel (C), where the storage time is $T_L/2$, $p_D$ undergoes a complete
revival. The energy of the retrieved signal field shown in panel (D) is therefore much larger, by a factor of $5.73$, than that in panel (B).
This is in good agreement with the DSP theory for retrieval efficiency discussed later, which predicts that the retrieved signal energy of panel
(D) should be $5.53$ times that in panel (B). These results demonstrate the importance of the adiabatic concept of DSP for a realistic
experimental scenario. The retrieved signal field directly reflects the DSP dynamics in the magnetic field.

We can predict the retrieval efficiency of a stored signal pulse by
tracking the population of the DSP as it evolves under the influence
of the magnetic field. To compute the polariton population, it is
convenient to consider the Fourier components of the DSP. We express
the polariton annihilation operator for the mode of wavenumber $q$
as
\begin{equation}
  \hat{\Psi}_{\alpha}\left(  q,t\right)
  = \frac{i\Omega(t)\hat{a}_{k,\alpha}-\sqrt{Np}\kappa ^{\ast}
          \sum_{m} R_{m\alpha}(\beta)
           \hat{S}_{g^{\prime}m+\alpha-\beta}^{gm}\left(  q,t\right)  }
    {\sqrt{ \Omega^{2} + Np \left\vert \kappa \right\vert ^{2}
    \sum_{m}
        R_{m\alpha}(\beta)^{2}}}%
\end{equation}
where $\hat{S}_{g^{\prime} m^{\prime}}^{gm} \equiv
(1/\sqrt{Np})\sum_{\mu} \hat{\sigma}_{gm, g^{\prime}
  m^{\prime}}^{(\mu)} \exp(-i(qz_{\mu} + (\omega_g -
\omega_{g^{\prime}})(t-z_{\mu}/c)))$ is a collective spin wave
annihilation operator of wavenumber $q$. These operators obey
quasi-bosonic commutation relations $
[\hat{S}_{g^{\prime}m_1^{\prime}}^{gm_1}(q),
  \hat{S}_{g^{\prime}m_2^{\prime}}^{gm_2}(q^{\prime})] =
\delta_{m_1 m_2} \delta_{m_1^{\prime}m_2^{\prime}} \delta_{q
  q^{\prime}} + O(1/N)$, and the DSP operators therefore also obey
 $[\Psi_a(q),\Psi_{a'}^{\dag}(q^{\prime})] =
\delta_{qq^\prime} \delta_{aa'}+O(1/N)$.
\begin{figure}
  \includegraphics{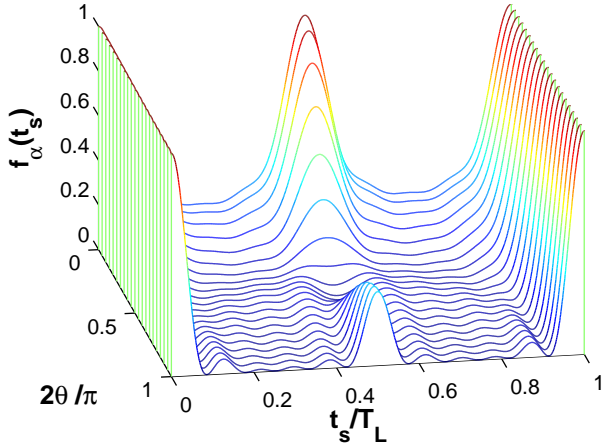}
  \caption{(Color online) The DSP population fraction $f_{\alpha}(t_s)$ calculated
  for orientations of the magnetic field $0 \le \theta \le \pi/2$ over
  one Larmor period.  These results illustrate collapses and revivals
  whose features are dependent on $\theta$.  The
  atomic configuration and field
  polarizations $\alpha$ and $\beta$ are described in the text.  }
  \label{FigRevival}
\end{figure}
During the storage, the evolution of the spin wave operators is given by $\hat{S}^{gm}_{m+\alpha-\beta}(q,t) = \sum_{m_1=-F_g}^{F_g}
\sum_{m_2=-F_{g'}}^{F_{g'}}\mathcal{D}^{(g)\dag}_{m_1 m} (t) \mathcal{D}^{(g')}_{m+\alpha-\beta,m_2} (t)\hat{S}^{gm_1}_{g'm_2}(q,0)$, where
$\mathcal{D}_{m,m^{\prime}}^{(s)}(t) \equiv \langle s,m\vert \exp(-ig_s \mathbf{\Omega}_B\cdot\mathbf{\hat{F}}t) \vert s,m^{\prime}\rangle$ is
the matrix element of the rotation operator for states in hyperfine level $s$, and $\mathbf{\Omega}_B \equiv \mu_B\mathbf{B}/\hbar$. Using the
bosonic commutation relations for  the spin wave operators, we can calculate the number of polaritons $\langle \hat{N_\alpha}(t_s) \rangle =
\langle\sum_q \hat{\Psi}^{\dag}_\alpha (q,t_s) \hat{\Psi}_\alpha(q,t_s)\rangle$ as a function of storage time $t_s$ for an arbitrary DSP quantum
state created in the storage process. In the limit of infinite control field amplitude, this converts into the total number of photons in the
retrieved signal field $\sum_k \hat{a}^{\dagger}_{k,\alpha} \hat{a}_{k,\alpha}$. We therefore derive an expression for the signal retrieval
efficiency as the fraction $f_\alpha(t_s) \equiv \langle \hat{N}_\alpha(t_s) \rangle/\langle \hat{N}_\alpha(0)\rangle$:
\begin{eqnarray}
  f_{\alpha}(t_s)
   &=& \Bigg\vert \sum_{m_1 m_2}\frac{R_{m_1 \alpha}(\beta) R_{m_2 \alpha}(\beta)}
       {\sum_m
       \left\vert R_{m\alpha}(\beta)\right\vert ^{2}}
       \mathcal{D}_{m_2 m_1}^{(g)  }(t_s)
            \nonumber\\
  &\times& \mathcal{D}_{m_1+\alpha-\beta, m_2+\alpha-\beta}^{(g^{\prime}) \dag}
                    \left(t_s\right)\Bigg\vert^2  .\label{EqPolPop}
\end{eqnarray}
In Fig. \ref{FigRevival}, we display the $f_{\alpha}(t_s)$ as a function of $t_s$ for a variety of magnetic field orientations. We again consider
an ensemble of $^{85}$Rb atoms with the same choice of atomic configuration and field polarizations discussed earlier. For $t_s \ll T_L$, we
observe a collapse in the polariton population, yielding an approximate retrieval efficiency of $f_{\alpha}(t_s) \approx \exp(-\eta_{\alpha}^2
(\Omega_L t_s)^2/2)$, where the collapse rate $\eta_{\alpha}$ depends on the angle, $\theta$, between the magnetic field and the propagation
axis. For $\theta=0$, we find
\begin{equation}
  \eta_{\alpha}^2(\theta=0) = 4\sum_{m_1,m_2}
                       \frac{|R_{m_1\alpha}(\beta)R_{m_2\alpha}(\beta)|^2}
                       {\left(\sum_m |R_{m\alpha}(\beta)|^2 \right)^2}
                       (m_1-m_2)^2.
\end{equation}
With the approximation $g_g = -g_{g^{\prime}}$, valid for ground level alkalis, it is clear that the system undergoes a revival to the initial
state after a complete Larmor period, and thus the signal retrieval efficiency should equal the zero storage time value. Depending on the
orientation of the magnetic field, we observe also a partial revival at half the Larmor period. For a magnetic field oriented along the $z$ axis,
the system dynamics are relatively simple. Each hyperfine coherence $\hat{S}_{g^{\prime} m+\alpha-\beta}^{g~m}$ merely picks up a phase factor
that oscillates at $m+(\alpha-\beta)/2$ times twice the Larmor frequency, thus returning the system to its initial state at half the Larmor
period. In this case, the partial revival is actually a full revival. On the other hand, for $\theta=\pi/2$, a rotation through half the Larmor
period causes the coherence transformation $\hat{S}_{g^{\prime}~m+\alpha - \beta}^{g~m} \rightarrow
\hat{S}_{g^{\prime}~-(m+\alpha-\beta)}^{g~-m}$ up to an overall phase factor. As a result, for the choice of equal field polarizations ($\alpha =
\beta$), the retrieval efficiency at half the Larmor period simplifies to $(\sum_m R_{m \alpha}(\alpha)R_{-m,\alpha}(\alpha) / \sum_m |R_{m
  \alpha}(\alpha)|^2)^2$, resulting in a partial revival.
For other orientations of the magnetic field, particularly for $\theta=\pi/4$, the revival at half the Larmor period is suppressed. This reflects
the more complicated dynamics of the individual spin coherences $\hat{S}_{g^{\prime}~m'}^{g~m}$ each of which transform into a superposition of
all $(2F_g+1)(2F_{g^{\prime}}+1)$ spin coherences, with complex time dependent coefficients governed by the rotation matrices. Stated physically,
there is a strong destructive interference between the various spin coherences when $\theta \approx \pi/4$.

We have developed a theory of the DSP as a mechanism to store and
retrieve light pulses in a degenerate unpolarized atomic medium.
The role of the DSP and its connection to storage retrieval
efficiency have been verified by full numerical solutions of the
propagation equations for a classical incident signal field. In
the presence of a magnetic field, we have demonstrated that the
DSP population undergoes collapses and revivals
during the pulse storage time. We predict that this polariton
dynamics is directly reflected in the signal pulse retrieval
efficiency. Our results may find applications in quantum
communication and computation approaches that utilize quantum
memories \cite{duan,briegel,knill,lim}.

This work was supported by NASA, Office of Naval Research, National
Science Foundation, Research Corporation, Alfred P. Sloan Foundation,
and Cullen-Peck Chair.

{\it Note added}: While this manuscript was being prepared, the
collapses and revivals were observed \cite{matsukevich2}, in
excellent agreement with the predictions of this work.

\end{document}